\documentclass[aps,prl,twocolumn,showpacs,preprintnumbers]{revtex4}
\usepackage{amsmath}
\usepackage{dcolumn}
\usepackage{bm}
\usepackage{subfigure}
\usepackage{amsfonts}
\usepackage{amssymb}
\usepackage{makeidx}
\usepackage{epsfig}
\usepackage{graphicx}

\begin{document}

\title{Collisionless kinetic regimes for quasi-stationary axisymmetric
accretion disc plasmas}
\author{C. Cremaschini$^{a}$ and M. Tessarotto$^{b}$}
\affiliation{$^{a}$International School for Advanced Studies (SISSA) and INFN, Trieste,
Italy\\
$^{b}$Department of Mathematics and Geosciences, University of Trieste,
Trieste, Italy}
\date{\today }

\begin{abstract}
This paper is concerned with the kinetic treatment of quasi-stationary
axisymmetric collisionless accretion disc plasmas. The conditions of
validity of the kinetic description for non-relativistic magnetized and
gravitationally-bound plasmas of this type are discussed. A classification
of the possible collisionless plasma regimes which can arise in these
systems is proposed, which can apply to accretion discs around both
stellar-mass compact objects and galactic-center black holes. Two different
classifications are determined, which are referred to respectively as
energy-based and magnetic field-based classifications. Different regimes are
pointed out for each plasma species, depending both on the relative
magnitudes of kinetic and potential energies and the magnitude of the
magnetic field. It is shown that in all cases, there can be quasi-stationary
Maxwellian-like solutions of the Vlasov equation. The perturbative approach
outlined here permits unique analytical determination of the functional form
for the distribution function consistent, in each kinetic regime, with the
explicit inclusion of finite Larmor radius-diamagnetic and/or
energy-correction effects.
\end{abstract}

\maketitle

\section{Introduction}

The dynamics of astrophysical accretion discs (ADs) still has aspects which
remain to be satisfactorily understood concerning the physical mechanisms
which are responsible for the occurrence of accretion flows. Historically,
treatments of accretion discs have often been made in terms of a purely
fluid-dynamical approach to which was added an \textquotedblleft
anomalous\textquotedblright\ form of viscosity (i.e. one not due to binary
particle collisions), following simple intuitive turbulence models such as
that giving rise to $\alpha $-discs \cite{king}. However, a widespread view
is that the origin of the effective viscosity lies in magnetic phenomena
(such as the magneto-rotational instability, MRI \cite{B91,B98}) and that
the medium needs to be treated as a magnetized plasma, when making detailed
investigations, rather than as a simple un-magnetized neutral fluid. Almost
always, these calculations are then performed within MHD treatments \cite%
{vietri}. When treating collisionless or weakly-collisional plasmas,
\textquotedblleft stand-alone\textquotedblright\ fluid descriptions
formulated independently of an underlying kinetic theory can usually
provide, at best, only a partial description of the plasma phenomenology,
and may therefore become inadequate or incorrect in this case. This is
because of a number of possible inconsistencies which may arise \cite{Cr2010}%
. Firstly, the MHD description does not generally include the correct
constitutive fluid equations for the fluid fields. In particular, the set of
fluid equations is generally not closed, thus requiring the independent
prescription of equations of state which may give rise to an incorrect
description if not based on kinetic theory. Furthermore, in fluid approaches
no account is usually given of microscopic phase-space particle dynamics
(including single-particle conservation laws) or of phase-space plasma
collective phenomena (kinetic effects). Kinetic effects may give rise to
plasma regimes in which the various particle species satisfy distinctive
asymptotic orderings. The proper identification of these kinetic regimes is
a necessary prerequisite for the development of consistent kinetic theory.
This implies, in particular, the adoption of multi-species kinetic
treatments, in contrast to single-species descriptions characteristic of
typical MHD approaches, such as the ideal-MHD model. These issues are
naturally addressed within a kinetic treatment, an approach which becomes
mandatory for collisionless plasmas. In fact, in the framework of kinetic
theory, all fluid fields are in principle consistently determined from the
kinetic distribution function (KDF) $f_{s}(y,t)$, with $y\equiv \left( 
\mathbf{r},\mathbf{v}\right) $, which describes the statistical properties
of the plasma at a microscopic level. Unlike stand-alone fluid approaches, a
fundamental feature of kinetic theory is the adoption of phase-space
techniques, which rely on perturbative expansions which typically hold only
on suitable subsets of velocity or phase spaces. A typical example of this
type is provided by gyrokinetic (GK) theory, which is generally applicable
only to magnetized particles (see discussion below). In this paper we intend
to stress that additional phase-space expansions are required for the proper
treatment of collisionless plasmas.

The application of this technique includes in principle both stationary and
quasi-stationary configurations (kinetic equilibria) as well as
dynamically-evolving AD plasmas, such as those subject to kinetic
instabilities. In this regard, the question arises concerning the physical
conditions under which kinetic equilibria can be realized and how they are
related to fluid or MHD treatments. Extending the work developed in Refs.%
\cite{Cr2010,Cr2011}, the goal of this investigation is to point out the
existence of a variety of possible collisionless regimes which may
characterize plasma species in axisymmetric ADs around compact objects. The
identification of these regimes is obtained by analyzing the physical
conditions for their realization in AD plasmas around both stellar-mass
compact objects and galactic black holes. In particular, in this paper the
problem is investigated of the existence of Maxwellian kinetic equilibria in
collisionless AD plasmas, providing their explicit analytical representation
relevant for AD phenomenology.

The results follow from a Vlasov-Maxwell kinetic description for
non-relativistic axisymmetric plasmas. It is demonstrated that
quasi-stationary Maxwellian-like kinetic equilibria exist which are
characterized by a number of notable features in the various regimes. These
include non-uniform fluid fields and differential azimuthal rotation,
temperature anisotropy and possibly quasi-stationary accretion flows not
dependent on turbulence phenomena. As pointed out in Ref.\cite{PRL},
collisionless plasmas can be stable under such general conditions. A case of
interest is represented by collisionless plasmas characterized by ion and
electron species having different temperatures, with typically $T_{e}\ll
T_{i}$. An example is provided by the radiatively inefficient flows (RIAFs)
arising in low-density geometrically-thick discs around black holes \cite%
{Narayan}. In these physical conditions, EM radiation effects\ on particle
dynamics, produced either by background radiation fields or
radiation-reaction phenomena \cite{EPJ2}, are negligible. Collisionless
plasmas can in principle consist of multiple ion and multiple electron
species (with indices $s=1,n$), each one being described by its velocity
KDF. In particular, each species carries individual characteristic times
associated with the Larmor rotation ($\tau _{Ls}$), the Langmuir time ($\tau
_{p}$) and the collision time ($\tau _{Cs}$), which can in principle be
determined from experimental observations. For definiteness, let us consider
AD plasmas having\ very different characteristic time scales, in the sense $%
\tau _{p},\tau _{Ls}\ll \Delta t\ll \tau _{Cs}$. Here, consistent with the
previous inequalities, $\Delta t=\Delta L/v_{the}$\ and $\Delta L$\ are the
equilibrium scales, namely the largest possible characteristic time and
length scales allowed for the fluid fields, with $v_{the}$\ being the
electron thermal velocity (see definition below). Plasmas satisfying these
orderings are referred to as collisionless and quasi-stationary, while at
the same time\ being characterized by a mean-field EM interaction and a
quasi-neutral charge density. In particular, we stress that
quasi-stationarity is intended here as slow time-variation with respect to
the time scales $\tau _{Ls}$\ and $\tau _{p}$. Within this framework, each
plasma species is described by a KDF which satisfies the Vlasov kinetic
equation $\frac{d}{dt}f_{s}(y,t)=0$, with the velocity moments determining
the system fluid fields and the sources of the EM self-fields $\left\{ 
\mathbf{E}^{self},\mathbf{B}^{self}\right\} $. In particular, ignoring
possible weakly-dissipative effects (Coulomb collisions and turbulence) and
instabilities (see for example Refs.\cite{Lomi09,Rebusco,Narayan2006}), this
paper focuses on regimes which are purely collisionless.

Two different criteria for the definition of kinetic regimes are introduced,
which are referred to as energy-based and magnetic field-based
classifications. The first one takes into account the relative magnitudes of
thermal and potential energies and leads to the identification of two
possible regimes, denoted respectively as strong and weak effective
potential energy regimes (SEPE and WEPE; see Section 4). The second one
instead depends on the magnitude of the magnetic field and gives rise to
four different possible kinetic regimes, here denoted respectively as
strongly-magnetized, intermediately-magnetized of type 1 and 2 and
weakly-magnetized plasma regimes (see Section 5). The physical conditions
characteristic of AD plasmas which give rise to these regimes are discussed.
Estimates of the order-of-magnitude of the magnetic field required for the
occurrence of each regime are also given, corresponding to physical
interesting situations occurring both in AD plasmas around stellar-mass
black holes and in galactic-center ADs.

\subsection{\textbf{Scheme of the paper}}

The paper is organized as follows. In Section 2 the basic assumptions of the
theory are presented. The necessary adiabatic invariants used for the
construction of kinetic equilibria are derived in Section 3, where the
relevant small dimensionless parameters for the regime classification are
also defined. Section 4 deals with the energy-based classification, leading
to the definition of the strong and weak effective potential energy regimes.
The magnetic field-based classification and the physical conditions for
existence of the different regimes in AD plasmas are presented in Section 5.
Section 6 describes the perturbative solution method adopted for the Vlasov
equation. Explicit construction of kinetic equilibria for the different
regimes identified is then addressed in Sections 7 and 8, where exact
Maxwellian-like solutions are obtained and their Chapman-Enskog
representations are given. Section 9 contains a discussion about the
construction of global solutions in mixed kinetic regimes. Comparison with
previous literature is given in Section 10, while Section 11 deals with the
explicit construction of the kinetic solution for each regime in a
particular example-case of astrophysical interest. Final concluding remarks
are presented in Section 12.\bigskip

\section{Basic assumptions}

We consider a non-relativistic multi-species AD plasma, in the sense that:
a) the gravitational field can be treated within classical Newtonian theory;
b) it has non--relativistic species flow velocities; c) all particles are
non-relativistic. Condition b) requires that for each plasma species the
related thermal velocities are non-relativistic. Condition c) implies in
turn the validity of b) and allows the non-relativistic Vlasov kinetic
equation to be used for describing collisionless plasmas.

Quasi-stationary solutions for the magnetic field $\mathbf{B}$ are
considered which admit a family of locally-nested axisymmetric toroidal
magnetic surfaces (which may be either locally open or closed \cite%
{Cr2010,Cr2011}), so that a set of magnetic coordinates ($\psi ,\varphi
,\vartheta $) can be prescribed locally in such a way that $\psi $\ is
identified with the poloidal flux, i.e. an observable, while $\vartheta $\
and $\varphi $\ are curvilinear angle-like coordinates defined on each
magnetic surface $\psi (\mathbf{x})=const.$ In particular, in this way the
value of $\psi $\ is uniquely determined at each point $x$\ once its value
on a reference flux surface is prescribed. As an example, $\psi (\mathbf{x}%
)=0$\ in the limit in which the corresponding magnetic surface reduces to a
circumference. Each relevant physical quantity can then be conveniently
expressed either in terms of the non-ignorable cylindrical coordinates $%
\mathbf{x\equiv }\left( R,z\right) $ or as a function of the corresponding
magnetic coordinates $\left( \psi ,\vartheta \right) $. The EM field is
taken of the form $\left[ \mathbf{E}(\mathbf{x},\lambda ^{k}t),\mathbf{B}(%
\mathbf{x},\lambda ^{k}t)\right] $, with $\lambda $ being a suitable small
dimensionless parameter and $k$ an integer $\geq 1$. In particular, the
magnetic field is $\mathbf{B}=\mathbf{B}^{self}(\mathbf{x},\lambda ^{k}t)+%
\mathbf{B}^{ext}(\mathbf{x},\lambda ^{k}t)$, where $\mathbf{B}^{self}$ and $%
\mathbf{B}^{ext}$ denote respectively the self-generated magnetic field
produced by the AD plasma, and a finite external magnetic field. The total
magnetic field $\mathbf{B}$ is then decomposed as $\mathbf{B}=\mathbf{B}_{T}+%
\mathbf{B}_{P}$, where $\mathbf{B}_{T}\equiv I(\mathbf{x},\lambda
^{k}t)\nabla \varphi $ and $\mathbf{B}_{P}\equiv \nabla \psi (\mathbf{x}%
,\lambda ^{k}t)\times \nabla \varphi $ are the toroidal and poloidal
components. Finally, it is also assumed that charged plasma particles are
subject to the action of the effective potential $\Phi _{s}^{eff}(\mathbf{x}%
,\lambda ^{k}t)=\Phi (\mathbf{x},\lambda ^{k}t)+\frac{M_{s}}{Z_{s}e}\Phi
_{G}(\mathbf{x},\lambda ^{k}t)$, with $\Phi (\mathbf{x},\lambda ^{k}t)$ and $%
\Phi _{G}(\mathbf{x},\lambda ^{k}t)$ denoting the corresponding
electrostatic (ES) and gravitational contributions, the latter generated
both by the compact object and the disc. The ES potential must be retained
in order to warrant the validity of the Poisson equation. In particular, the
origin of the electric field can be ascribed to the requirement of
satisfying quasi-neutrality in the presence of differentially-rotating AD
plasmas. Since the ES field is uniquely prescribed by the Poisson equation,
it follows that the ideal Ohm's law, typically used in ideal-MHD, may not
hold anymore for collisionless plasmas.

A final remark concerns the physical meaning of the potentials $\Phi $\ and $%
\Phi _{G}$\ which enter the definition of $\Phi _{s}^{eff}$. Provided both $%
\Phi _{G}$\ and $\left\vert \nabla \Phi _{G}\right\vert $\ vanish in the
limit in which $\left\vert \mathbf{x}\right\vert \rightarrow \infty $, $\Phi
_{G}$\ is a unique solution of the Poisson-mass equation $\nabla ^{2}\Phi
_{G}=4\pi G\rho _{m}$, with $\rho _{m}$\ denoting the mass-density.
Similarly, quasi-neutrality determines uniquely $\Phi $. Therefore, both $%
\Phi $\ and $\Phi _{G}$\ must be intended as observables.

\section{Adiabatic invariants and dimensionless parameters}

The construction of quasi-stationary kinetic solutions requires the
identification of the relevant dynamical invariants, i.e., first integrals
of motion or more generally adiabatic invariants characterizing
single-particle dynamics. Because of axisymmetry the toroidal canonical
momentum $p_{\varphi s}$ is a first integral of motion. This is given by%
\begin{equation}
p_{\varphi s}=M_{s}R\mathbf{v}\cdot \mathbf{e}_{\varphi }+\frac{Z_{s}e}{c}%
\psi \equiv \frac{Z_{s}e}{c}\psi _{\ast s},
\end{equation}%
where $M_{s}$ is the particle mass and $\mathbf{e}_{\varphi }$ is a unit
vector along the azimuthal direction $\varphi $. Thanks to the assumption
introduced for the EM fields, the total particle energy%
\begin{equation}
E_{s}=\frac{M_{s}}{2}v^{2}\mathbf{+}{Z_{s}e}\Phi _{s}^{eff}\equiv {Z_{s}e}%
\Phi _{\ast s}
\end{equation}%
is an adiabatic invariant. Possible additional adiabatic invariants can be
determined provided GK theory applies. In this case dynamical variables are
evaluated at the particle guiding-center position (and here labeled with a
prime). In particular, the particle magnetic moment $m_{s}^{\prime }$ and
the guiding-center canonical momentum\textit{\ }$p_{\varphi s}^{\prime
}\equiv \frac{M_{s}}{B^{\prime }}\left( u^{\prime }I^{\prime }-\frac{c\nabla
^{\prime }\psi ^{\prime }\cdot \nabla ^{\prime }\Phi _{s}^{^{\prime }eff}}{%
B^{\prime }}\right) +\frac{Z_{s}e}{c}\psi ^{\prime }$ are useful GK
invariants, with $u^{\prime }$ denoting the component of the guiding-center
particle velocity parallel to the local direction of the magnetic field.

In order to classify the plasma regimes and the conditions providing the
corresponding quasi-stationary kinetic solutions, we introduce the
dimensionless species parameters $\varepsilon _{M,s}$, $\varepsilon _{s}$
and $\sigma _{s}$. It is convenient to prescribe them in such a way to be
all independent of single-particle velocity and at the same time to be
related to the characteristic species thermal velocities. In general, for
this purpose both perpendicular and parallel thermal velocities (defined
with respect to the magnetic field direction) must be considered. These are
defined respectively by $v_{\perp ths}=\left\{ T_{\perp s}/M_{s}\right\}
^{1/2}$\ and $v_{\parallel ths}=\left\{ T_{\parallel s}/M_{s}\right\} ^{1/2}$%
, with $T_{\perp s}$\ and $T_{\parallel s}$\ denoting the species
perpendicular and parallel temperatures. In particular, the first parameter
is defined as $\varepsilon _{M,s}\equiv \frac{r_{Ls}}{L}$, where $%
r_{Ls}=v_{\perp ths}/\Omega _{cs}$\ is the species average Larmor radius,
with $\Omega _{cs}=Z_{s}eB/M_{s}c$\ being the species Larmor frequency. Here 
$L$\ is the minimum scale-length characterizing the spatial variations of
all of the fluid fields associated with the KDF, and of the EM fields. The
second parameter $\varepsilon _{s}$\ is related to the particle canonical
momentum $p_{\varphi s}$. Denoting by $v_{ths}\equiv \sup \left\{
v_{\parallel ths},v_{\perp ths}\right\} $, $\varepsilon _{s}$\ is identified
with $\varepsilon _{s}\equiv \left\vert \frac{M_{s}Rv_{ths}}{\frac{Z_{s}e}{c}%
\psi }\right\vert $. Hence, $\varepsilon _{s}$\ effectively measures the
ratio between the toroidal angular momentum $L_{\varphi s}\equiv
M_{s}Rv_{\varphi }$\ and the magnetic contribution to the toroidal canonical
momentum, for all particles in which $v_{\varphi }$\ is of the order $%
v_{\varphi }\sim v_{ths}$\ while $\psi $\ is assumed as being non-vanishing.
In particular, here the magnetic flux can be estimated as $\psi \sim
B_{p}RL_{1}$, with $L_{1}$\ denoting the characteristic length-scale of flux
variations and $B_{p}$ the magnitude of the poloidal magnetic field. Note
that, by definition, $L\leq L_{1}$\ and in principle $L$\ can be $\ll L_{1}$%
\ locally. Finally, $\sigma _{s}$\ is related to the particle total energy $%
E_{s}$\ and is prescribed as $\sigma _{s}\equiv \left\vert \frac{\frac{M_{s}%
}{2}v_{ths}^{2}}{{Z_{s}e}\Phi _{s}^{eff}}\right\vert $. It follows that $%
\sigma _{s}$\ measures the ratio between particle kinetic and potential
energies, for all particles having velocity $v$\ of the order $v\sim v_{ths}$%
, with $\Phi _{s}^{eff}$ being assumed as non-vanishing. In the following we
shall denote as thermal subset of velocity space the subset of the Euclidean
velocity space in which the asymptotic conditions $\frac{v}{v_{ths}}\sim 
\frac{v_{\varphi }}{v_{ths}}\sim O\left( 1\right) $ holds.

\section{Plasma regimes I: energy-based classification}

Different classifications can be introduced for collisionless AD plasmas
according to the magnitude of the three independent parameters $\sigma _{s}$%
, $\varepsilon _{s}$ and $\varepsilon _{M,s}$ defined above. The
classification associated with the parameter $\sigma _{s}$ is referred to as
the \textit{energy-based classification}, while $\varepsilon _{s}$ and $%
\varepsilon _{M,s}$ determine the \textit{magnetic field-based classification%
}. In this section we consider the energy-based classification. Each species
in the AD plasma is said to be in the regimes of\newline
A) \emph{Strong effective potential energy (SEPE regime) }if $\sigma _{s}\ll
1$,\newline
B) \emph{Weak effective potential energy (WEPE regime) }if $\sigma
_{s}\lesssim 1$,

if the respective inequalities are satisfied. We stress that in both cases
all particles characterized by velocities of the order $v\lesssim v_{ths}$\
are considered as being gravitationally and/or electrostatically bound,
namely confined in a finite sub-set of the Euclidean configuration space.
Such a condition is manifestly a physical prerequisite for the existence of
AD systems.

In the case of regime A) the following asymptotic expansion holds for the
total particle energy $\Phi _{\ast s}$:%
\begin{equation}
\Phi _{\ast s}\mathbf{=}\Phi _{s}^{eff}\left[ 1+O\left( \sigma _{s}\right) %
\right] .  \label{ordering-sigma}
\end{equation}%
Let us briefly discuss the possible physical mechanisms which may be
responsible for the establishment of these regimes. It is obvious that the
SEPE regime requires the action of some energy non-conserving mechanism. Two
limiting cases can be considered for this regime: when $Z_{s}e\Phi
_{s}^{eff}(\mathbf{x},\lambda ^{k}t)\cong M_{s}\Phi _{G}(\mathbf{x},\lambda
^{k}t)$ and\ when $\Phi _{s}^{eff}(\mathbf{x},\lambda ^{k}t)\cong \Phi (%
\mathbf{x},\lambda ^{k}t)$ respectively. In the first case, plausible
physical mechanisms that can be responsible for the decrease of the
single-particle kinetic energy, in both collisionless and collisional AD
plasmas, are EM interactions (e.g., binary Coulomb collisions among
particles and particle-wave interactions, such as Landau damping) and/or
radiation emission (radiation-reaction). In particular, EM interactions can
in principle be ascribed also to the occurrence of EM instabilities and EM
turbulence. For single particles these processes can be dissipative, i.e.
can involve the loss of kinetic energy. As a consequence, these particles
tend to move towards regions with higher gravitational potential (in
absolute value). After multiple interactions of this type, the process can
ultimately give rise to the SEPE regime. In the second case, it is assumed
that quasi-neutrality can locally generate a strong ES potential. The
strength of this potential depends in turn on the charge density produced by
the plasma species. In rotating plasmas the latter is primarily affected by
the centrifugal and gravitational potentials and the poloidal magnetic flux.
On the other hand, the WEPE regime corresponds to configurations having
comparable kinetic and effective potential energies.

\section{Plasma regimes II: magnetic field-based classification}

Next, we address the magnetic field-based classification with respect to the
parameters $\varepsilon _{s}$ and $\varepsilon _{M,s}$. Here, plasma species
can be distinguished as being in the following asymptotic regimes:\newline
1) \emph{Strongly-magnetized} if $\varepsilon _{M,s}\ll 1$ and $\varepsilon
_{s}\ll 1$.\newline
2) \emph{Intermediately-magnetized of type 1 }if $\varepsilon _{M,s}\ll 1$
but $\varepsilon _{s}\sim 1$.\newline
3) \emph{Intermediately-magnetized of type 2 }if $\varepsilon _{M,s}\sim 1$
but $\varepsilon _{s}\ll 1$.\newline
4) \emph{Weakly-magnetized} if $\varepsilon _{M,s}\sim 1$ and $\varepsilon
_{s}\gtrsim 1$.

GK theory applies only when $\varepsilon _{M,s}\ll 1$, namely for species
belonging to cases 1) and 2). Furthermore, when the ordering $\varepsilon
_{s}\ll 1$ holds, the following asymptotic expansion holds for the particle
canonical momentum $\psi _{\ast s}$:%
\begin{equation}
\psi _{\ast s}=\psi \left[ 1+O\left( \varepsilon _{s}\right) \right] .
\label{ordering-epsilon}
\end{equation}

From the physical point of view, the magnetic-based classification can be
justified as follows. We first notice that the magnetic field affects the
two parameters $\varepsilon _{s}$ and $\varepsilon _{M,s}$ in different
ways. In the first case it enters by means of the poloidal flux $\psi $
which contributes to the toroidal canonical momentum $p_{\varphi s}$, while
in the second case what matters is the magnitude of the total magnetic
field. Invoking the definitions given above for $\varepsilon _{s}$ and $%
\varepsilon _{M,s}$\ it follows that $\varepsilon _{s}\sim \varepsilon _{M,s}%
\frac{L}{L_{1}}\frac{B}{B_{p}}$, where $L$\ and $L_{1}$\ are respectively
the minimum scale-lengths of equilibrium fluid and EM fields and of the
poloidal flux. In general, the two quantities should be considered as
independent, with $L\leq L_{1}$ and $B_{p}\leq B$.\ Indeed, the parameter $%
\varepsilon _{s}$ determines the particle spatial excursions from a magnetic
flux surface, while $\varepsilon _{M,s}$\ measures the amplitude of the
Larmor radius with respect to the inhomogeneities of the background fluid
fields. These two effects correspond to two different physical
magnetic-related processes, due respectively to the Larmor-radius and
magnetic-flux surface confinement mechanisms. This justifies the
magnetic-field based classification given above. In particular, case 1)
holds when the Larmor radius remains small with respect to the minimum
scale-length $L$\ and, at the same time, the particle trajectory remains
close to the same magnetic surface $\psi =const.$\ Case 2) applies when, in
difference to case 1), the departure of particle trajectories from the $\psi 
$-surfaces becomes non-negligible in comparison with $L_{1}$. This can occur
in the case in which $B_{p}\ll B\frac{L}{L_{1}}$, which can happen only when
a strong toroidal magnetic field is present. Case 3) instead arises when the
Larmor radius becomes comparable to $L$, while the particle azimuthal
angular momentum remains much smaller than the magnetic part of the
corresponding canonical momentum. This happens only when $L_{1}\gg L\frac{B}{%
B_{p}}$, a situation which may occur, for example, when the EM field is
primarily externally generated. It must be noted that, in this case, the
poloidal flux varies on the largest scale $L_{1}$, so that variations of $%
\psi $ occurring on the Larmor radius scale are negligible under this
condition. Finally, case 4) arises when particle trajectories undergo finite
excursions from both the magnetic field lines and the $\psi $-surfaces.

The magnetic-based classification provided here affords interesting
applications to the physics of AD plasmas. In this regard, a link between
the asymptotic regimes and observed astrophysical objects is needed. We
consider first the case of strongly-magnetized hydrogen-ion plasma species ($%
s=i$), and estimate the minimum magnitude of the magnetic field for which
this regime can occur. For simplicity one can consider $L\sim L_{1}$ and $%
v_{\varphi }\sim v_{thi}$, which gives $\varepsilon _{M,s}\sim \varepsilon
_{s}\ll 1$. Let us consider two representative examples of stellar-mass and
galactic-center mass black holes. In the first case we take $M_{\ast }\sim
10M_{\odot }$ (giving a Schwarzschild radius $R_{Sch}\sim 30km$) as
representative of the black hole mass and consider plasma located at a
distance $R\sim 10-100\,R_{Sch}$ from the central object, with ion
temperatures in the range $T_{i}\sim 10^{4}-10^{11}K$, and with
characteristic scale-length $L\sim 1-10R_{Sch}$. Then, requiring $%
\varepsilon _{M,i}\lesssim 10^{-j}$, with $j\geq 1$, we get that $B\gtrsim
10^{j-1}\,G$ for the highest temperature and smallest $L$, and $B\gtrsim
10^{j-6}\,G$ for the lowest temperature and largest $L$. For example,
setting $\varepsilon _{M,i}\lesssim 0.01$ requires $j=2$ in these estimates.
For a galactic-center black hole: taking mass $M_{\ast }=10^{8}M_{\odot }$,
the equivalent estimates, for the same range of radial distances in terms of
Schwarzschild radii, give $B\gtrsim 10^{j-10}\,G$ and $B\gtrsim 10^{j-14}\,G$
respectively for the two sets of parameters. Next, consider an example of
regime 2), taking $L\sim L_{1}$ and $v_{\varphi }\gg v_{thi}$. In this case
the previous estimates for $B$ from $\varepsilon _{M,i}$ remain unchanged,
while $\varepsilon _{s}\sim 1$ requires $\frac{v_{thi}}{v_{\varphi }}\sim
\varepsilon _{M,i}$, i.e. that the ion species is supersonic. The
corresponding estimates of the minimum value of $B$ required for regimes 3)
and 4) in the case of galactic-center black holes give extremely low values.
This indicates that these regimes are unlikely in such a case, and could
only be relevant for ADs around stellar-mass black holes. In particular, for
regime 3), taking $v_{\varphi }\sim v_{thi}$ and $L\ll L_{1}$ gives $%
\varepsilon _{M,i}\sim 1$ and $\varepsilon _{i}\ll 1$ when $B\sim 10^{-1}G$.
Instead, for regime 4), taking again $v_{\varphi }\sim v_{thi}$ but $L\sim
L_{1}$, one obtains that $\varepsilon _{M,i}\sim 1$ for $B\sim 10^{-6}$. We
conclude that, in practice, the majority of collisionless AD plasmas around
galactic-center black holes are actually expected to belong to regime 1),
while in the case of stellar-mass black holes all regimes could occur.

As a final point, note that the classification defined here completely
departs from the one usually adopted in MHD treatments based on the
one-fluid description, which involves the specification of the magnitude of
the dimensionless parameter $\beta \equiv \frac{8\pi p}{B^{2}}$. Here, as
usual, $p$ denotes the isotropic thermal pressure of the plasma. The two
classifications are indeed intrinsically different because they concern
single-particle dynamics and single-fluid dynamics respectively. It is easy
to show, on the other hand, that the requirement for strongly-magnetized
plasma species introduced above does not rule out at all any of the
possibilities of having high ($\beta \gg 1$), finite ($\beta \sim 1$) or low
($\beta \ll 1$) beta plasmas. In the particular case in which the toroidal
magnetic field is negligible, a connection can be established between $%
\varepsilon _{s}$ and $\beta $ by taking $v_{\varphi }\sim v_{thi}$ and $%
\psi \sim BRL_{1}$ as order-of-magnitude estimates. The following
relationship is obtained:%
\begin{equation}
\beta =8\pi \frac{n_{i}L_{1}^{2}}{M_{i}}\left( \frac{Z_{i}e}{c}\right)
^{2}\varepsilon _{i}^{2}=\frac{L_{1}^{2}}{\lambda _{D}^{2}}\frac{v_{thi}^{2}%
}{c^{2}}\varepsilon _{i}^{2}
\end{equation}%
where $n_{i}$ denotes the ion number density and $\lambda _{D}$ the
Debye-length. Considering the case of a hydrogen-ion plasma, with $L\sim
L_{1}$ and number density in the range $n_{i}\sim 10^{6}-10^{14}cm^{-3}$:
for the case of stellar-mass black holes one obtains the estimate of $\beta $
in the range $\beta \sim \left[ 10^{4}-10^{14}\right] \varepsilon _{i}^{2}$.
Then, assuming for example $\varepsilon _{i}$ in the interval $\varepsilon
_{i}\sim 10^{-8}-10$, it follows that depending on the magnitude of $%
\varepsilon _{i}$, all of the ranges of $\beta $ indicated above are in
principle permitted. Analogous estimates can be obtained also for
galactic-center black holes, giving in all cases $\beta \gg 1$ when $%
\varepsilon _{i}$ is in the same interval. Therefore, the high-beta regime
appears more likely to occur in this case. The conclusion is therefore that
the kinetic treatment considered here encompasses all of the regimes for $%
\beta $ usually considered in MHD treatments of AD plasmas.

\bigskip

\section{Solution \textbf{method}}

Concerning the method adopted for constructing the solution of the Vlasov
equation, we follow here the perturbative theory developed in Refs.\cite%
{Cr2010,Cr2011,Cr2011a}. For each plasma species, the KDF is represented as
an expansion in terms of a complete set of functions. The latter ones can
always be identified with suitable generalized Gaussian distributions. For a
collisionless plasma, this is equivalent to effectively decomposing the
system in terms of particle sub-species. In principle, two approaches are
possible for determining the sub-species KDFs. The first one is based on the
Chapman-Enskog solution of the Vlasov equation by seeking a perturbative
solution of the form $f_{s}=f_{M,s}+\lambda f_{1s}+...$,\ where $f_{M,s}$\
is a drifted Maxwellian KDF and $\lambda $ is a suitable dimensionless small
parameter. However, this approach does not generally take into account
\textquotedblleft a priori\textquotedblright\ the exact conservation laws of
particle dynamics. In the case of a magnetized plasma the latter should also
include conservation of the corresponding GK invariants. The construction of
the Chapman-Enskog solution requires the determination of the perturbations $%
\lambda ^{k}f_{ks}$, for $k=1,2,...$, which involves explicitly solving
appropriate PDEs. An alternative approach, which avoids this difficulty, is
to construct an exact (or asymptotic) solution of the Vlasov equation of the
form $f_{s}=f_{\ast s}$, where $f_{\ast s}$\ is a suitable adiabatic
invariant, so that it is necessarily a function of all of the independent
particle adiabatic invariants. This technique has been developed
systematically in Refs.\cite{Cr2010,Cr2011,Cr2011a} and is the one also
adopted here. In particular, in the following we show that, depending on
which of the different possible regimes identified above is being
considered, particular solutions of the Vlasov equation of the second type
can be consistently obtained. In all regimes $f_{\ast s}$\ is proved to be
asymptotically \textquotedblleft close\textquotedblright\ (in a suitable
sense)\ to either a local Maxwellian or bi-Maxwellian KDF. The advantage of
this method is that it also permits determining \textquotedblleft a
posteriori\textquotedblright\ a perturbative representation of the KDF
equivalent to the Chapman-Enskog expansion, which consistently retains
finite Larmor-radius (FLR), diamagnetic and/or energy corrections to the
KDF. This can be achieved, for each kinetic regime, by implementing the
appropriate Taylor expansions with respect to the dimensionless parameters $%
\sigma _{s}$ and $\varepsilon _{s}$.

It is understood that the basic feature of such a kinetic perturbative
technique is that it is only strictly applicable in localized subsets of
velocity space (thermal subsets), namely to particles whose velocity
satisfies the asymptotic ordering (\ref{ordering-sigma}) and/or (\ref%
{ordering-epsilon}). A notable consequence of such an approach is that, for
each kinetic regime, quasi-stationary, self-consistent, asymptotic solutions
of the Vlasov-Maxwell equations (kinetic equilibria) can be explicitly
determined by means of suitable Taylor expansions of $f_{\ast s}$. In
particular, it is found that Maxwellian-like KDFs can be obtained locally in
phase-space, where the appropriate convergence conditions hold. This
procedure provides also the correct constitutive equations of the
leading-order fluid fields as well as the precise form of the
FLR-diamagnetic and energy-correction contributions to the KDF.

\section{Quasi-stationary solutions for plasmas in the SEPE regime}

We now prove explicitly the existence of quasi-stationary kinetic solutions
for each of the four magnetic-based regimes holding in the limit of $\sigma
_{s}\ll 1$. We consider first the case of strongly-magnetized and type 1
intermediately-magnetized plasmas. Since $\varepsilon _{M,s}\ll 1$, the
species quasi-stationary KDFs can be expressed in terms of exact and GK
adiabatic invariants. In both regimes 1) and 2) the KDF is taken to be of
the form%
\begin{equation}
f_{\ast s}=f_{\ast s}\left( E_{s},\psi _{\ast s},p_{\varphi s}^{\prime
},m_{s}^{\prime },\Lambda _{\ast s},\lambda ^{k}t\right) ,  \label{form}
\end{equation}%
where $\Lambda _{\ast s}$ denotes the so-called structure functions, i.e.,
functions which depend implicitly on the particle state $\mathbf{x}$ and
must be properly prescribed according to the specific form of the solution
(see below). For definiteness, both $f_{\ast s}$ and $\Lambda _{\ast s}$ are
assumed to be analytic functions in terms respectively of $\Lambda _{\ast s}$
and $\mathbf{x}$. In order for $f_{\ast s}$ to be an adiabatic invariant, $%
\Lambda _{\ast s}$ must also be a function of the adiabatic invariants. This
restriction is referred to as a kinetic constraint. The two regimes 1) and
2) differ in the precise functional dependences imposed on $\Lambda _{\ast
s} $, which are taken to be of the type $\Lambda _{\ast s}=\Lambda
_{s}\left( \Phi _{\ast s},\psi _{\ast s}\right) $ and $\Lambda _{\ast
s}=\Lambda _{s}\left( \Phi _{\ast s}\right) $ for strongly-magnetized and
type 1 intermediately-magnetized regimes respectively. Invoking Eqs.(\ref%
{ordering-sigma}) and (\ref{ordering-epsilon}) for $\Lambda _{\ast s}$, it
follows that the structure functions can be Taylor-expanded in the two cases
to give 
\begin{eqnarray}
\Lambda _{\ast s} &=&\Lambda _{s}\left( \Phi _{s}^{eff},\psi \right) \left[
1+O\left( \varepsilon _{s}\right) +O\left( \sigma _{s}\right) \right] , \\
\Lambda _{\ast s} &=&\Lambda _{s}\left( \Phi _{s}^{eff}\right) \left[
1+O\left( \sigma _{s}\right) \right] .
\end{eqnarray}%
The actual definition of the structure functions and of the corresponding
kinetic constraints follows by identifying the quantities $\Lambda
_{s}\left( \Phi _{s}^{eff},\psi \right) $ and $\Lambda _{s}\left( \Phi
_{s}^{eff}\right) $ with appropriate sets of fluid fields, namely velocity
moments of the KDF $f_{\ast s}$. In turn, these choices depend on the form
of $f_{\ast s}$. An example consistent with the previous requirements is
given by a generalized bi-Maxwellian KDF, which in the notation of Refs.\cite%
{Cr2010,Cr2011} is given by%
\begin{eqnarray}
f_{\ast s} &=&\frac{\widehat{\beta _{\ast s}}}{\left( 2\pi /M_{s}\right)
^{3/2}\left( T_{\parallel \ast s}\right) ^{1/2}}  \notag \\
&&\times \exp \left\{ -\frac{H_{\ast s}}{T_{\parallel \ast s}}+\frac{%
p_{\varphi s}^{\prime }\xi _{\ast s}}{T_{\parallel \ast s}}-m_{s}^{\prime }%
\widehat{\alpha _{\ast s}}\right\} .  \label{sol2}
\end{eqnarray}%
Here $H_{\ast s}\equiv E_{s}-\frac{Z_{s}e}{c}\psi _{\ast s}\Omega _{\ast s}$
and $\widehat{\alpha _{\ast s}}\equiv \frac{B^{\prime }}{\widehat{\Delta
_{T_{s}}}}$, with the quantities $\frac{1}{\widehat{\Delta _{T_{s}}}}\equiv 
\frac{1}{\widehat{T}_{\perp s}}-\frac{1}{T_{\parallel \ast s}}$, $\widehat{T}%
_{\perp s}$, $T_{\parallel \ast s}$ being associated with the species
temperature anisotropy and the perpendicular and parallel temperatures
respectively. As a consequence, the structure functions are identified in
this case with the set $\left\{ \Lambda _{\ast s}\right\} \equiv \left\{ 
\widehat{\beta _{\ast s}},\widehat{\alpha _{\ast s}},T_{\parallel \ast
s},\Omega _{\ast s},\xi _{\ast s}\right\} $, where $\widehat{\beta _{\ast s}}
$ is related to the species number density, and $\Omega _{\ast s}$ and $\xi
_{\ast s}$ are related to the azimuthal and the poloidal velocities. By
construction, Eq.(\ref{sol2}) is an asymptotic solution of the Vlasov
equation, in terms of which the fluid fields are uniquely determined. It
follows that the related velocity moment equations are also identically
satisfied. Due to the smoothness assumption, a Chapman-Enskog representation
of Eq.(\ref{sol2}) can be recovered by applying to the structure functions,
when appropriate, the Taylor expansions (\ref{ordering-sigma}) and (\ref%
{ordering-epsilon}), which hold in suitable subsets of velocity space. This
perturbative solution gives a formal representation of the KDF of the type%
\begin{equation}
f_{\ast s}=f_{bi-M,s}\left[ 1+\varepsilon _{s}h_{Ds}^{\left( 1\right)
}+\sigma _{s}h_{Ds}^{\left( 2\right) }\right] ,  \label{kkk}
\end{equation}%
which holds in the thermal subset of velocity space where the SEPE regime
applies, so that $v\sim v_{\varphi }\sim v_{ths}$\ with $\sigma _{s}\ll 1$.
Depending whether $\varepsilon _{s}$ is either $\varepsilon _{s}\ll 1$\ or $%
\varepsilon _{s}\sim 1$, the $\varepsilon _{s}$-expansion is applicable or
not. In the second case, namely for intermediately-magnetized species of
type 1, the perturbative correction $h_{Ds}^{\left( 1\right) }$\ is
effectively null. Here the notation is as follows. First,\textbf{\ }$%
f_{bi-M,s}$ denotes the leading-order contribution, which coincides with a
drifted bi-Maxwellian KDF carrying non-uniform number density, and azimuthal
and poloidal flow velocities, as well as temperature anisotropy. Hence, the
bi-Maxwellian KDF $f_{bi-M,s}$\ should be considered itself an asymptotic
solution. In addition, the functional dependences in terms of $\left( \Phi
_{s}^{eff},\psi \right) $ or $\Phi _{s}^{eff}$ (cases 1) and 2)
respectively) remain arbitrary. Second, $h_{Ds}^{\left( 1\right) }$ and $%
h_{Ds}^{\left( 2\right) }$ identify the first-order FLR-diamagnetic and
energy-correction terms respectively. By construction, to leading-order in
the expansion parameters, $h_{Ds}^{\left( 1\right) }$ and $h_{Ds}^{\left(
2\right) }$ are polynomial functions of the particle velocity which depend
linearly on the so-called thermodynamic forces, namely the gradients $\frac{%
\partial \Lambda _{s}}{\partial \psi }$ and/or $\frac{\partial \Lambda _{s}}{%
\partial \Phi _{s}^{eff}}$. For the specific calculations of the
perturbative contributions $h_{Ds}^{\left( 1\right) }$ and $h_{Ds}^{\left(
2\right) }$ we refer to paper \cite{Cr2011} and related discussion.

We now analyze the case of intermediately-magnetized plasmas of type 2 and
weakly-magnetized plasmas. In both cases the general solution of the Vlasov
equation cannot depend on GK invariants and therefore is necessarily of the
form%
\begin{equation}
f_{\ast s}=f_{\ast s}\left( E_{s},p_{\varphi s},\Lambda _{\ast s},\lambda
^{k}t\right) ,  \label{solweak}
\end{equation}%
with $k\geq 1$ and $f_{\ast s}$ and $\Lambda _{\ast s}$ again assumed to be
analytic functions. As for the previous cases, regimes 3) and 4) differ in
the precise functional dependences imposed on $\Lambda _{\ast s}$, here
assumed to be of the types $\Lambda _{\ast s}=\Lambda _{s}\left( \Phi _{\ast
s},\psi _{\ast s}\right) $ and $\Lambda _{\ast s}=\Lambda _{s}\left( \Phi
_{\ast s}\right) $ respectively. A possible form is given in terms of the
species generalized drifted Maxwellian KDF of the form%
\begin{equation}
f_{\ast s}=\frac{\eta _{\ast s}}{\left( 2\pi /M_{s}\right) ^{3/2}T_{\ast
s}^{3/2}}\exp \left\{ -\frac{E_{s}-\Omega _{\ast s}p_{\varphi s}}{T_{\ast s}}%
\right\} .  \label{weakk}
\end{equation}%
Here the structure functions are $\Lambda _{\ast s}\equiv \left( \eta _{\ast
s},T_{\ast s},\Omega _{\ast s}\right) $, where $\eta _{\ast s}$, $T_{\ast s}$
and $\Omega _{\ast s}$ are related to the species number density, isotropic
temperature and azimuthal angular velocity respectively. A perturbative
Taylor expansion of Eq.(\ref{weakk}) obtained invoking Eqs.(\ref%
{ordering-sigma}) and (\ref{ordering-epsilon}), leads to the Chapman-Enskog
representation%
\begin{equation}
f_{\ast s}=f_{M,s}\left[ 1+\varepsilon _{s}h_{Ds}^{\left( 3\right) }+\sigma
_{s}h_{Ds}^{\left( 4\right) }\right] ,  \label{pppp}
\end{equation}%
where the leading-order contribution $f_{M,s}$ coincides with a drifted
isotropic Maxwellian KDF carrying non-uniform number density, azimuthal
differential flow velocity and isotropic temperature. Again, the asymptotic
representation (\ref{pppp}) holds in the thermal subset of velocity space
and in validity of the corresponding regimes for $\sigma _{s}$ and $%
\varepsilon _{s}$. The kinetic constraints require that the latter are
smooth functions either of $\left( \Phi _{s}^{eff},\psi \right) $ (regime
3)) or of $\Phi _{s}^{eff}$ (regime 4)). Instead, the perturbative
first-order corrections $h_{Ds}^{\left( 3\right) }$ and $h_{Ds}^{\left(
4\right) }$ are polynomial functions of the particle velocity, with $%
h_{Ds}^{\left( 3\right) }=0$ for weakly-magnetized plasmas, which contain
again FLR-diamagnetic and energy-correction contributions through the
thermodynamic forces $\frac{\partial \Lambda _{s}}{\partial \Phi _{s}^{eff}}$
and/or $\frac{\partial \Lambda _{s}}{\partial \Phi _{s}^{eff}}$. The
detailed expressions for the perturbative contributions $h_{Ds}^{\left(
3\right) }$ and $h_{Ds}^{\left( 4\right) }$ can be obtained based on the
technique outlined in Ref.\cite{Cr2011} (see also Section 10 for a specific
application of this approach).

\bigskip

\section{Quasi-stationary solutions for plasmas in the WEPE regime}

We now address the issue of\ the existence of quasi-stationary kinetic
solutions for each of the four regimes identified by the magnetic
field-based classification, but now considering the limit of the WEPE
regime. In particular, when $\sigma _{s}\lesssim 1$ the asymptotic expansion
given by Eq.(\ref{ordering-sigma}) cannot apply. This restriction strongly
affects the physical realizability of kinetic equilibria in this regime. We
consider first the case of strongly-magnetized plasmas. In this case GK
theory can be formulated and the quasi-stationary KDF can be assumed to be
of the general form expressed by Eq.(\ref{form}). A convenient
representation can then still be given in terms of a generalized
bi-Maxwellian KDF as indicated in Eq.(\ref{sol2}). For strongly-magnetized
plasmas, the WEPE regime differs from the corresponding SEPE regime by the
functional dependences imposed on the structure functions $\Lambda _{\ast s}$%
. In order to warrant the existence of Maxwellian-like kinetic solutions,
the only admissible form for the kinetic constraint is of the type $\Lambda
_{\ast s}=\Lambda _{s}\left( \psi _{\ast s}\right) $. Hence, invoking Eq.(%
\ref{ordering-epsilon}) for $\Lambda _{\ast s}$, it follows that the
structure functions can be Taylor-expanded to give%
\begin{equation}
\Lambda _{\ast s}=\Lambda _{s}\left( \psi \right) \left[ 1+O\left(
\varepsilon _{s}\right) \right] .  \label{asympt-wepe1}
\end{equation}%
This in turn implies for the KDF $f_{\ast s}$,\textbf{\ }in the thermal
subset of velocity space, the asymptotic expansion%
\begin{equation}
f_{\ast s}=f_{bi-M,s}\left[ 1+\varepsilon _{s}h_{Ds}^{\left( 5\right) }%
\right] ,  \label{mmm}
\end{equation}%
where again the leading-order contribution $f_{bi-M,s}$ coincides with a
drifted bi-Maxwellian KDF characterized by non-uniform number density,
azimuthal and poloidal flow velocities, as well as temperature anisotropy,
and whose functional dependences in terms of $\psi $ remain arbitrary. It
should be noted that in the WEPE regime the perturbative correction $%
h_{Ds}^{\left( 5\right) }$ can only contain contributions due to first-order
FLR-diamagnetic terms. It follows immediately that, to leading-order in the
expansion parameter $\varepsilon _{s}$, the KDF $h_{Ds}^{\left( 5\right) }$
is a polynomial function of the particle velocity which depends linearly on
the gradients $\frac{\partial \Lambda _{s}}{\partial \psi }$.

A similar analysis can be carried out for the regime 3) for species
belonging to intermediately-magnetized plasmas of type 2. In this case the
GK adiabatic invariants do not exist, so that the quasi-stationary KDF must
be of the type defined by Eq.(\ref{solweak}). This can be satisfied, in
particular, by the species generalized drifted Maxwellian KDF given in Eq.(%
\ref{weakk}). Also in this case, the existence of a Maxwellian-like
equilibrium solution requires imposing a kinetic constraint of the type $%
\Lambda _{\ast s}=\Lambda _{s}\left( \psi _{\ast s}\right) $, which implies
again the validity of the asymptotic expansion in Eq.(\ref{asympt-wepe1}).
When applied to the KDF this provides the Chapman-Enskog representation%
\begin{equation}
f_{\ast s}=f_{M,s}\left[ 1+\varepsilon _{s}h_{Ds}^{\left( 6\right) }\right] ,
\end{equation}%
which is again applicable in the thermal subset of velocity space. Here, the
leading-order contribution $f_{M,s}$ coincides with a drifted isotropic
Maxwellian KDF carrying non-uniform number density, azimuthal differential
flow velocity and isotropic temperature. In this case the structure
functions are found to be smooth functions of the poloidal flux $\psi $.
Instead, to leading-order in $\varepsilon _{s}$, the perturbative correction 
$h_{Ds}^{\left( 6\right) }$ again contains FLR-diamagnetic contributions and
depends linearly on the thermodynamic forces $\frac{\partial \Lambda _{s}}{%
\partial \psi }$. We again stress here that the calculation of the
contributions $h_{Ds}^{\left( 5\right) }$ and $h_{Ds}^{\left( 6\right) }$
follows from the perturbative theory developed in Ref.\cite{Cr2011}. An
illustration of the method is also presented in Section 10.

An important remark concerns the validity of the two WEPE regimes in which
all species admit the $\varepsilon _{s}-$expansion. For definiteness, let us
consider the case of a two-species hydrogen ion-electron plasma. As proved
in Ref.\cite{Cr2011a}, if the ordering assumption $\Omega _{i}R\sim v_{thi}$
is invoked (with $\Omega _{i}$ denoting the ion azimuthal rotation
frequency), quasi-neutrality necessarily implies that the ES potential must
satisfy the ordering $\left\vert \frac{e\Phi }{T_{i}}\right\vert \sim
1/O\left( \varepsilon _{i}\right) $, while $\left\vert \frac{\frac{e}{c}%
\Omega _{i}\psi }{T_{i}}\right\vert \sim 1/O\left( \varepsilon _{i}\right) $%
. As a consequence, $O\left( \sigma _{i}\right) \sim O\left( \varepsilon
_{i}\right) \ll 1$, which violates the initial assumption $\sigma
_{s}\lesssim 1$. To restore the consistency of the WEPE orderings, it must
therefore be required that\ $\left\vert \frac{\frac{e}{c}\Omega _{i}\psi }{%
T_{i}}\right\vert \sim O(1)$, implying $\frac{\Omega _{i}R}{v_{thi}}\sim
O\left( \varepsilon _{i}\right) $.

A separate analysis must be performed in the WEPE regime for the two
magnetic field-based classifications for species belonging to
intermediately-magnetized plasmas of type 1\emph{\ }and weakly-magnetized
plasmas. For these regimes, both of the asymptotic expansions (\ref%
{ordering-sigma}) and (\ref{ordering-epsilon}) remain forbidden. An
equilibrium solution in terms of generalized bi-Maxwellian and Maxwellian
KDFs can be obtained in both cases. On the other hand, the requirement of
recovering at the same time a Chapman-Enskog representation of the solution
which warrants the existence of Maxwellian-like equilibria necessarily
implies that all of the structure functions are either identically constant,
namely $\Lambda _{\ast s}=const.$, or contain suitably-slow dependences with
respect to the variables $\left( \mathbf{r},t\right) $. In other words, in
the latter case there should exist a dimensionless small parameter $\delta
\ll 1$ so that the $\Lambda _{\ast s}$ are still adiabatic invariants of the
form $\Lambda _{\ast s}=\Lambda _{s}\left( \delta ^{n}\mathbf{r},\delta
^{n}t\right) $, with $n\geq 1$. Such solutions correspond to either a
spatially-uniform and rigid-rotating plasma or to a slowly-varying one.
Although mathematically admissible, both of these are generally not
acceptable from the physical point of view, unless the AD plasma is
characterized by constant or slowly-varying fluid fields in the sense
indicated above. These conditions may fail, for example, near to the
boundaries. We therefore conclude that, in the WEPE regime,
intermediately-magnetized plasmas of type 1\emph{\ }and weakly-magnetized
plasmas may not admit physically-acceptable Maxwellian-like equilibria.

\bigskip

\section{Global solution for mixed regimes}

Mixed regimes can occur when particles belonging to the thermal subset of
velocity space can move between mutually accessible spatial domains
corresponding to different kinetic regimes. The question arises of the very
existence of Maxwellian-like equilibria in these cases and how the
equilibrium KDF can be obtained. A positive answer can be reached only
provided in each separate regime being considered, a Maxwellian-like
solution exists and at same time at least one of the two asymptotic
parameters $\varepsilon _{s}$\ and $\sigma _{s}$\ remains $\ll 1$\ in all
mixed regimes. In such a case the global equilibrium KDF is always
determined by Taylor expansion with respect to the small parameter common to
all the mixed regimes, based on the adoption of the perturbative technique
pointed out in the previous sections. This solution method in principle
excludes the possibility of having mixed regimes in the presence of
intermediately-magnetized plasmas of type 1\ and weakly-magnetized plasmas
in the WEPE regime.

We discuss here specific examples in which only two adjacent regions A and B
of configuration space are responsible for the occurrence of mixed regimes.
For this purpose, it suffices to treat the following three examples. The
first case is illustrated by the situation in which A is in the
strongly-magnetized SEPE regime while B is in the strongly-magnetized WEPE
regime. Therefore, the two domains differ only for the magnitude of the
parameter $\sigma _{s}$, which is respectively $\ll 1$ in A and $\sim 1$ in
B. In this case the equilibrium KDF is determined uniquely by the solution
corresponding to the strongly-magnetized WEPE regime and is given by Eq.(\ref%
{mmm}). It follows that the structure functions can only be of the form $%
\Lambda _{\ast s}=\Lambda _{s}\left( \psi _{\ast s}\right) $. Therefore, the
global generalized bi-Maxwellian solution $f_{\ast s}$ generates to
first-order only FLR-diamagnetic corrections, while energy-correction terms
due to the effective potential expansion are ruled out.

The second case of interest is the one in which both regions A and B are in
the SEPE regime, with A being strongly-magnetized and B
intermediately-magnetized of type 1. In this mixed regime the global
solution is determined by the generalized bi-Maxwellian given by Eq.(\ref%
{sol2}) with the structure functions allowed to depend only on the total
particle energy, namely they are of the form $\Lambda _{\ast s}=\Lambda
_{s}\left( \Phi _{\ast s}\right) $. Hence, in this configuration, only
energy corrections appear in the global solution, which therefore becomes of
the form of Eq.(\ref{kkk}) with $h_{Ds}^{\left( 1\right) }=0$.

Finally, the third example is provided by domains A and B which are both in
the SEPE regime, with A being strongly-magnetized and B
intermediately-magnetized of type 2 respectively. In this configuration the
global solution $f_{\ast s}$ cannot depend on GK adiabatic invariants, while
the structure functions remain of the general form $\Lambda _{\ast
s}=\Lambda _{s}\left( \Phi _{\ast s},\psi _{\ast s}\right) $. It follows
that the global solution is given by the generalized drifted Maxwellian KDF
in Eq.(\ref{weakk}), which admits the asymptotic Chapman-Enskog
representation corresponding to Eq.(\ref{pppp}).

\bigskip

\section{Comparison with the literature}

Let us now address the issue of comparison of the present work with the
relevant previous literature. The comparison here is limited only to studies
dealing with kinetic treatments of axisymmetric plasmas. The majority of
these have considered kinetic equilibria for laboratory plasmas and are not
in practice applicable to AD plasmas. In fact, ADs are intrinsically
different, at least because of: a) the physical contexts; b) the effective
potential acting on plasma particles, which depends on both the ES and the
gravitational potentials; c) the topology of the magnetic flux lines; d) the
asymptotic orderings, which are generally quite different for laboratory and
astrophysical plasmas; e) the kinetic boundary conditions holding in the two
cases. In contrast with laboratory plasmas, the gravitational field plays a
fundamental role in determining both the equilibrium solutions and their
stability properties \cite{Cr2011,PRL}. An analysis of the relevant
literature for laboratory plasmas is nevertheless useful in order to
understand possible connections and clarify the conclusions drawn here. We
first consider a number of papers belonging to this category.

Historically, a first comparison of this type can be made with the so-called
astron equilibria \cite{Love79}. These are characterized by non-Maxwellian
equilibrium KDFs. Such distributions are intended as implicit functions of
single particle energy and canonical momentum, which can only describe
rigidly-rotating ring plasmas. This configuration is uninteresting for
realistic AD plasmas having a finite extension in the configuration domain.
Similar conclusions can be drawn for the theory of Vlasov equilibria
developed for laboratory field-reversed plasmas presented in Ref.\cite%
{chan80}, where again non-Maxwellian ring-plasmas were considered. Kinetic
equilibria of various types can also be found in several papers dealing with
linear stability analysis. An earlier example case is provided by Ref.\cite%
{Spar84}, where axisymmetric plasma rings were treated in terms of
rigidly-rotating Maxwellian equilibria. Another example is provided by Ref.%
\cite{Morse87}, which makes use of a Chapman-Enskog solution method to
determine the equilibrium KDF for toroidal plasmas. Also in this case,
however, the leading-order KDF is identified with a rigidly-rotating
isotropic Maxwellian. In Ref.\cite{Wong91}, instead, analogous annular
configurations have been described by means of monoenergetic Dirac-delta
KDFs. In general, besides these features which are manifestly incompatible
with AD systems, all these studies ignore the existence of stationary
electric fields as well as of GK adiabatic invariants, and they are not
suited for the description of differentially-rotating AD plasmas
characterized by shear-flow in the presence of a gravitational field. A
notable work in which differential rotation in laboratory toroidal plasmas
has been consistently dealt with is that due to Catto \textit{et al.} \cite%
{Catto1987}. The equilibrium KDF in this case was expressed in terms of the
canonical momentum and total kinetic energy, yielding a generalized
isotropic Maxwellian equilibrium consistent with the Chapman-Enskog
representation. As a basic consequence, it was found that the constitutive
equations for the fluid fields were uniquely determined and subject to
specific kinetic constraints. Recently, the approach has been generalized in
Ref.\cite{Cr2011a} to the treatment of axisymmetric rotating Tokamak plasmas
in the collisionless regime. This work takes into account in a consistent
way the constraints imposed by single-particle conservation laws as well as
those imposed by the Maxwell equations. The solution obtained allows one to
describe Tokamak plasmas which are generally characterized by equilibrium
azimuthal and poloidal differential rotations, non-uniform fluid fields and
temperature anisotropy.

Let us now briefly summarize some relevant contributions specific to AD
plasmas. The first work to be mention is the one due to Bhaskaran and
Krishan \cite{Mahajan01}, based in turn on the theoretical approach
developed by Mahajan \cite{Mahajan02, Mahajan03} for laboratory plasmas. A
Chapman-Enskog solution method is implemented. This allows one to represent
the equilibrium KDF in terms of an infinite power series in terms of the
ratio of the drift velocity to the thermal speed (considered as the small
expansion parameter). To leading-order, this recovers a spatially
homogeneous Maxwellian distribution. Therefore, the physical applicability
of the approach remains strongly limited (see also the discussion below).

More recent investigations concern the adoption of kinetic closure
conditions in MHD numerical simulations. This issue is particularly relevant
for stability investigations of collisionless AD plasmas (see for example
Refs.\cite{Quataert2002,Sharma2006,Quataert2007B, Snyder1997}). These\ kind
of studies are based on fluid equations which are coupled to suitable
kinetic closure conditions. However, they rely on single-fluid descriptions
based on the ideal Ohm's law and, in addition the single-species kinetic
equilibrium is usually identified either with a Maxwellian or a
bi-Maxwellian KDF having uniform number density and temperature, but
otherwise exhibiting a differential azimuthal rotation. As we show in the
next section, the theory developed in this paper allows one to take into
account more general kinetic equilibria, which hopefully afford a more
consistent treatment of plasma phenomenology occurring in actual ADs.

\bigskip

\section{Example case}

We now show how the formalism outlined in the previous sections can be
implemented in practice to determine explicitly consistent kinetic
equilibria for the different regimes identified. The discussion is also
useful in order to establish a deeper comparison with previous literature,
such as Refs.\cite{Quataert2002,Quataert2007B}.

We adopt cylindrical coordinates $\left( R,\varphi ,z\right) $, with $%
\varphi $ still representing the ignorable coordinate. We consider a disc
composed of a multi-species collisionless plasma subject to both
gravitational and EM fields. In particular, it is assumed that the
self-gravitational field is negligible in comparison with that generated by
the central object. The latter is expressed in terms of the Newtonian
potential associated with the central mass, which is of the form $\Phi
_{G}=\Phi _{G}\left( R,z\right) $. Concerning the magnetic field, the
external component is assumed to give the dominant contribution, while the
self-generated field is neglected. In particular, the magnetic field is
assumed to have uniform vertical and azimuthal components. In the notation
introduced here, the equilibrium magnetic field can be written as $\mathbf{B}%
\cong \mathbf{B}^{ext}=I\nabla \varphi +B_{z}\mathbf{e}_{z}$, with $B_{z}=%
\frac{1}{R}\frac{\partial \psi }{\partial R}$. This implies that the
poloidal magnetic flux function $\psi $ is necessarily of the form $\psi
=\alpha R^{2}$, with $\alpha $ being a suitable\ real constant. Therefore
the flux surfaces $\psi =const.$ coincide with vertical planes in an $R-z$
section of the disc at constant $\varphi $. Finally, the ES field is assumed
to be purely self-generated and of the form $\Phi =\Phi \left( R,z\right) $.
These requirements pose non-trivial constraints on the existence of kinetic
equilibria of this type (and in particular of Maxwellian equilibria), which
arise from the solubility conditions of the Poisson and the Ampere equations.

For definiteness, let us consider a species Maxwellian equilibrium KDF
characterized, at leading order in the relevant asymptotic parameters, by
isotropic temperature and purely azimuthal flow velocity of the form $%
\mathbf{V}_{s}=\Omega _{s}R\mathbf{e}_{\varphi }$, with the angular
frequency being generally of the type $\Omega _{s}=\Omega _{s}\left(
R,z\right) $. This kind of dependence for $\Omega _{s}$ is compatible, for
example, with a Keplerian angular frequency. In the literature using this
approximation, the KDF is often assumed to carry uniform species
temperatures. This choice is also adopted here for the sake of comparison.
This type of model is typically used for the stability analysis of AD
plasmas with respect to the magneto-rotational instability (MRI). We now
analyze whether these requirements can be satisfied in the various kinetic
regimes indicated above by a KDF of the form given by Eq.(\ref{solweak}). It
follows immediately that this can be achieved for all magnetic field-based
configurations belonging to either the SEPE or WEPE regimes. A
representation for the KDF $f_{\ast s}$ is then provided in fact by the
generalized Maxwellian distribution defined in Eq.(\ref{weakk}) with $%
T_{\ast s}$ being identified with the leading-order isotropic species
temperature $T_{s}=const.$, while the remaining structure functions $\Lambda
_{\ast s}$ are identified with the set $\Lambda _{\ast s}=\left( \eta _{\ast
s},\Omega _{\ast s}\right) $. An equivalent representation of $f_{\ast s}$
follows using the definitions for $E_{s}$ and $p_{\varphi s}$, giving%
\begin{equation}
f_{\ast s}=\frac{n_{\ast s}}{\left( 2\pi /M_{s}\right) ^{3/2}T_{s}^{3/2}}%
\exp \left\{ -\frac{M_{s}\left( \mathbf{v}-\mathbf{V}_{\ast s}\right) ^{2}}{%
2T_{s}}\right\} ,  \label{maxwellll}
\end{equation}%
where $\mathbf{V}_{\ast s}\equiv \Omega _{\ast s}R^{2}\nabla \varphi $ and
the quantity $n_{\ast s}$ is defined as%
\begin{equation}
n_{\ast s}\equiv \eta _{\ast s}\exp \left[ \frac{\frac{M_{s}}{2}\Omega
_{\ast s}^{2}R^{2}-Z_{s}e\Phi _{s}^{eff}+\frac{Z_{s}e}{c}\psi \Omega _{\ast
s}}{T_{s}}\right] .
\end{equation}%
The validity of Maxwellian-like equilibria of this type is warranted for
each of the regimes identified above by imposition of the corresponding
appropriate kinetic constraints on $\Lambda _{\ast s}$\textbf{\ }(see the
discussion in the previous sections). Depending on the kinetic regimes being
considered, the KDF can be Taylor-expanded in terms of the dimensionless
parameters $\varepsilon _{s}$\ and/or $\sigma _{s}$\ and then represented in
terms of the leading-order structure functions $\Lambda _{s}=\left( \eta
_{s},\Omega _{s}\right) $. Consider first the SEPE regime. There, as a
fundamental consequence, for the example configuration considered here, the
structure functions are either of the form\ $\Lambda _{s}=\Lambda _{s}\left(
\Phi _{s}^{eff},\psi \right) $, for strongly-magnetized plasmas and type 2
intermediately-magnetized plasmas, or of the form $\Lambda _{s}=\Lambda
_{s}\left( \Phi _{s}^{eff}\right) $ for weakly and type 1
intermediately-magnetized plasmas. Next, we consider the WEPE regime. In the
two configurations identified above which admit the $\varepsilon _{s}$%
-expansion, one finds that necessarily $\Lambda _{s}=\Lambda _{s}\left( \psi
\left( R\right) \right) $. This means that $z$-dependences remain excluded
for these cases for the equilibrium KDF. Finally, in the weakly and type 1
intermediately-magnetized plasmas belonging to the WEPE regime, existence of
asymptotic equilibria can only be obtained by imposing the slow-dependence
condition $\Lambda _{\ast s}=\Lambda _{s}\left( \delta ^{n}\mathbf{r},\delta
^{n}t\right) $. This can be obtained, for example, by identifying the
infinitesimal parameter $\delta $\ with $\delta =\frac{r}{R}\ll 1$, with $r$
denoting a spatial displacement.

In summary, for each of the regimes considered above, the leading-order KDF
is obtained from Eq.(\ref{maxwellll}) by replacing $\left( n_{\ast s},%
\mathbf{V}_{\ast s}\right) $\ by $\left( n_{s},\mathbf{V}_{s}\right) $\
subject to the corresponding functional dependences pointed out here. The
KDF obtained in this way coincides with an isotropic drifted Maxwellian
distribution. In particular, it follows that the leading-order number
density $n_{s}$\ takes the form%
\begin{equation}
n_{s}\equiv \eta _{s}\exp \left[ \frac{\frac{M_{s}}{2}\Omega
_{s}^{2}R^{2}-Z_{s}e\Phi _{s}^{eff}+\frac{Z_{s}e}{c}\psi \Omega _{s}}{T_{s}}%
\right] .
\end{equation}

Excluding now the two WEPE regimes indicated above for which neither
Larmor-radius nor energy expansions are allowed, all of the other regimes
are characterized by non-vanishing diamagnetic and/or energy corrections to
the equilibrium KDF. These contributions come from the Taylor expansion of $%
f_{\ast s}$ and differ according to the specific regime considered. For an
illustration of the perturbative approach adopted here, we report explicitly
the calculations of the perturbative corrections $h_{Ds}^{\left( i\right) }$%
, $i=1,6$ corresponding to the sample case considered in this section. In
the absence of GK adiabatic invariants and within the validity of the
assumptions introduced, it follows that the $\varepsilon _{s}$-expansion
yields the formally analogous functions $h_{Ds}^{\left( 1\right)
}=h_{Ds}^{\left( 3\right) }=h_{Ds}^{\left( 5\right) }=h_{Ds}^{\left(
6\right) }$, while the $\sigma _{s}$-expansion gives similarly $%
h_{Ds}^{\left( 2\right) }=h_{Ds}^{\left( 4\right) }$. As a consequence, the
first-order correction terms in the two cases are found to be%
\begin{eqnarray}
h_{Ds}^{\left( 1\right) } &=&\frac{cM_{s}R}{Z_{s}e}\left[ \frac{\partial \ln
\eta _{s}}{\partial \psi }+\frac{p_{\varphi s}\Omega _{s}}{T_{s}}\frac{%
\partial \ln \Omega _{s}}{\partial \psi }\right] \mathbf{v}\cdot \mathbf{e}%
_{\varphi },  \label{a} \\
h_{Ds}^{\left( 2\right) } &=&\frac{M_{s}}{2{Z_{s}e}}\left[ \frac{\partial
\ln \eta _{s}}{\partial \Phi _{s}^{eff}}+\frac{p_{\varphi s}\Omega _{s}}{%
T_{s}}\frac{\partial \ln \Omega _{s}}{\partial \Phi _{s}^{eff}}\right] v^{2},
\label{bb}
\end{eqnarray}%
with $\eta _{s}$ and $\Omega _{s}$ being prescribed according to the kinetic
regimes indicated above. These terms generally imply non-vanishing
contributions to the relevant equilibrium fluid fields, namely the total
species number density, azimuthal flow velocity and isotropic temperatures.

From this analysis of the sample case it follows that a general form of the
equilibrium KDF is obtained such that:

1) For all of the plasma kinetic regimes (and hence independently of the
strength of the magnetic field and of the effective potential energy), to
leading-order the species KDF coincides with a Maxwellian equilibrium,
which, to leading-order, has uniform temperature.

2) Such equilibria are generally however not exact and require, for each
regime, consistent determination of the appropriate perturbative corrections
to the Maxwellian KDF, as mentioned here.

Nevertheless, the existence of these equilibria is subject to the validity
of the Maxwell equations, in particular quasi-neutrality and Ampere's
equation. The related discussion for strongly-magnetized plasmas in the SEPE
regime is given in Ref.\cite{Cr2011}. The analysis can be extended in
principle to all of the other kinetic regimes considered here.

These conclusions permit us to perform a comparison with the literature.
First it must be noted that the classification of AD plasmas usually
adopted, based on the $\beta -$parameter (see definition above), does not
rule out the existence of the kinetic regimes pointed out here for the
various plasma species. In particular, as shown above, the requirement $%
\beta \gg 1$ can in principle correspond to both SEPE and WEPE regimes as
well as to any of the magnetic field-based regimes defined in Sections 6 and
7.

An important point concerns the possibility of imposing, to leading-order,
uniform species number densities. This is clearly not permitted in the WEPE
regime, because $\Phi _{s}^{eff}$ is always a function of both $R$ and $z$
and this contradicts the functional form of $\eta _{s}$ required by the
kinetic constraints in that case. Instead, in principle this constraint
might still be satisfied in the SEPE regime by suitably prescribing the
coefficient $\eta _{s}$ according to the kinetic constraints. However,
quasi-neutrality in this case implies the vanishing of the ES potential. It
follows that for all species, the effective potential must coincide with the
gravitational potential, namely\ $Z_{s}e\Phi _{s}^{eff}=M_{s}\Phi _{G}$, so
that validity of the SEPE ordering requires $T_{s}/M_{s}\Phi _{G}\sim
O\left( \sigma _{s}\right) $. From analysis of the electron linear momentum
equation it follows however that this ordering condition cannot be satisfied
because $\Phi _{G}$ is a function of both $R$ and $z$, and so consistent
kinetic equilibria generally require a non-uniform species number density
and consequently also a non-uniform ES potential. This conclusion poses
serious limits on the possibility for realizing an equilibrium of this type
(i.e., with $n_{s}=const.$).

A further fundamental consequence of the kinetic treatment developed here is
that non-vanishing diamagnetic and/or energy correction terms to the
Maxwellian KDF may actually appear in several kinetic regimes. These
contributions are generally non-negligible and so they should be retained
consistently for both analytical and numerical treatments of these
equilibria. On the other hand, the specific functional form of these
corrections depends on the specific kinetic regime. As a further key
element, this means that in all cases prescription of the appropriate regime
is required. In turn, this implies that, for multi-species collisionless
plasmas, a multi-species treatment is generally required. Apart from
fundamental physical reasons, one obvious motivation for this is that
different plasma species can in principle belong to different kinetic
regimes.

To close this section, we should point out that the validity of these
conclusions is assured also in the case in which the equilibrium toroidal
magnetic field vanishes or remains negligible with respect to the poloidal
component. In those cases, if all plasma species belong to the
strongly-magnetized SEPE regime, the stability analysis given in Ref.\cite%
{PRL} applies. We therefore conclude that, even for the simplified model
considered in this section, the classification of the kinetic regimes
matters, and the perturbative kinetic theory developed here should be
invoked for both equilibrium and stability analysis.

\bigskip

\section{Conclusions}

In this paper we have proposed a classification of the species plasma
kinetic regimes which characterize collisionless accretion disc plasmas
around compact objects. The investigation has been based on non-relativistic
kinetic theory and has been carried out in the framework of a Vlasov-Maxwell
description. The case of collisionless axisymmetric magnetized plasmas has
been considered, for which the influence of radiation phenomena on
single-particle dynamics is negligible. It has been demonstrated that in all
of the regimes identified here, quasi-stationary Maxwellian-like kinetic
solutions exist. In particular, it has been shown that for each separate
kinetic regime or for suitably-mixed regimes, the quasi-stationary species
KDF can be uniquely obtained and represented in terms of generalized
Maxwellian or bi-Maxwellian distribution functions. A notable feature of the
approach is that the functional form of the species equilibrium KDF and the
constitutive equations for the leading-order fluid fields are determined
analytically by means of suitable perturbative expansions presented here.
This approach allows one to uniquely determine the first-order perturbative
contributions to the distribution function, which consistently retain all of
the relevant kinetic effects associated with the FLR-diamagnetic and
energy-correction terms. The procedure leads to a Chapman-Enskog-type
solution in which the leading-order term is identified with either a drifted
Maxwellian or a bi-Maxwellian distribution. The following important features
should be mentioned. The first one is that, independent of the strength of
the magnetic field, the local Maxwellian or bi-Maxwellian KDFs are generally
only approximate kinetic solutions of the Vlasov equation. The second
feature is that the perturbative theory can be developed in principle to
arbitrary order, thus also permitting analytic determination of the
corresponding fluid fields and moment equations to the requisite accuracy.
The inherent simplicity and clarity of the kinetic approach outlined here
provides the starting point for systematic kinetic stability analysis \cite%
{PRL} as well as for collisional \cite{Catto1987,White} and anomalous
transport theory. These features are relevant for theoretical and numerical
investigations of the phenomenology of AD plasmas.


\textbf{Acknowledgments - }This work has been partly developed within the
framework of MIUR (Italian Ministry for Universities and Research) PRIN
Research Programs and the GAMAS GRDE (Groupe des Recherces Europeene, CNRS,
France, Paris). The authors are indebted to J. C. Miller (Department of
Physics, University of Oxford, Oxford, UK) for his helpful comments and
suggestions. Stimulating discussions with John Papaloizou (Department of
Applied Mathematics and Theoretical Physics, Cambridge University,
Cambridge, UK), Ramesh Narayan (Harvard-Smithsonian Center for Astrophysics,
Harvard University, Cambridge, MA, USA), Gregory Hammett (Princeton Plasma
Physics Laboratory, Princeton University, Princeton, NJ, USA), Anatoly
Spitkovsky (Department of Astrophysical Sciences, Princeton University,
Princeton, NJ, USA) and Andrew MacFadyen (Department of Physics, New York
University, New York, NY, USA) are acknowledged by C.C.

\end{document}